\documentclass[a4paper,11pt]{article}
\pdfoutput=1 % if your are submitting a pdflatex (i.e. if you have
\usepackage{jcappub} % for details on the use of the package, please
\usepackage[T1]{fontenc} % if needed

\usepackage{xcolor}

\title{\boldmath A search for the fine-structure constant evolution from fast radio bursts and type Ia supernovae data}

\author[a,1]{Thais Lemos,\note{Corresponding author}}
\author[b,a]{Rodrigo Gon\c{c}alves,}
\author[a]{Joel Carvalho,}
\author[a]{Jailson Alcaniz}

\affiliation[a]{Observat\'orio Nacional, Rio de Janeiro - RJ, 20921-400, Brazil}
\affiliation[b]{ Departamento de F\'{\i}sica, Universidade Federal Rural do Rio de Janeiro, Serop\'edica - RJ, 23897-000, Brazil}
\emailAdd{thaislemos@on.br}
\emailAdd{rsg\_goncalves@ufrrj.br}
\emailAdd{jcarvalho@on.br}
\emailAdd{alcaniz@on.br}

\abstract{The search for a space-time variation of the fundamental constants has been explored over the years to test our physical theories. In this paper, we use the dispersion measure ($\mathrm{DM}$) of fast radio bursts (FRB) combined with type Ia supernovae (SNe) data to investigate a possible redshift evolution of the fine-structure constant ($\alpha$), considering the runaway dilaton scenario, which predicts  $\frac{\Delta \alpha}{\alpha} = - \gamma\ln{(1+z)}$, where $\gamma$ is a constant proportional to the current value of the coupling between the dilaton field and hadronic matter. We derive all the relevant expressions for the $\mathrm{DM}$ dependence concerning the fine-structure constant and constrain the parameter $\gamma$ from measurements of 17 well-localized FRBs and 1048 SNe data from the Pantheon compilation. We also use Monte Carlo simulations to forecast the constraining power of larger samples of FRB measurements for data sets with $N = 500$ and $N = 1000$ points. We found that the uncertainty on $\gamma$ can be improved by one order of magnitude and that limits on $\frac{\Delta \alpha}{\alpha}$ beyond $\sigma \sim 10^{-2}$ will depend crucially on better control of statistical and systematic uncertainties of upcoming FRB data.}

\begin{document}
\maketitle
\flushbottom

\section{Introduction}

Dirac's Large Numbers Hypothesis~\cite{Dirac1937} marked the initial exploration of the constancy of the fundamental constants of physics. This hypothesis suggested that the state of the Universe could influence dimensionless universal constants. In this context, investigations on the space-time evolution of such constants are crucial in physics, as any deviation from our current physical theories could indicate the presence of new physics~\cite{Uzan2003,Uzan2011}.

The time evolution of the fundamental constants could arise from different physical mechanisms, such as the existence of new particles of dark matter, magnetic monopoles or fundamental strings (see \cite{Martins2017} for a recent review). From the cosmological point of view, although most of the cosmological models assume the constancy of the physics fundamental constants,  there are several studies on the evolution of such constants, e.g., variation of the speed of light $c$ \cite{Magueijo2003}, the gravitational constant $G$ \cite{Gaztanaga2001}, the proton-to-electron mass ratio $\mu$ \cite{Safronova2018}, among others \cite{Martins2017}. 

The present analysis investigates a possible time dependence of the fine-structure constant, $\alpha$. Besides the critical implications on fundamental physics, this possibility has recently been suggested as a way to alleviate some problems in cosmology, such as the Hubble tension  \cite{Hart2020,Hart2021,Seto2023} and the primordial helium-4 abundance tension \cite{Seto2023_2} (see \cite{Martins2021,Deal2021} to discussion of time variation of $\alpha$ with other Big Bang Nucleosynthesis observations). A non-constant $\alpha$ has also several observational consequences related to the possible non-conservation of the number of photons along the geodesic, leading to a direct violation of the cosmic distance duality relation \cite{Minazzoli2014,Hess2014} and a deviation from a black body spectrum of the cosmic microwave background (CMB) radiation  \cite{Ade2015,Smith2019}.

Most theories in which the fine-structure constant presents a time dependence involve a scalar field controlling such dependence \cite{Dvali2002,Chiba2002,Martins2014}. In the present paper, we consider one of these models, the so-called runaway dilaton model, in which the dilaton plays the role of the scalar field responsible for coupling, yielding a time evolution of $\alpha$ \cite{Damour2002,Damour2002_2}. At low and intermediate redshifts, such evolution is given by
\begin{equation}
\label{eq:RunModel}
\frac{\Delta \alpha}{\alpha} = - \gamma\ln{(1+z)},
\end{equation}
where the constant parameter $\gamma$ is the product of the current value of the coupling between the dilaton and hadronic matter and the variation of the dilaton field with respect to the scale factor at the present time~(see e.g. \cite{Holanda:2015oda,Martins2015}).

We use observations of Fast Radio Bursts (FRBs) to find novel constraints on $\gamma$. FRBs are a millisecond transient event in radio frequency ($\sim$GHz) (for a review, see \cite{Petroff2022,Thornton2013,Petroff2015,Petroff2016,Platts2019}) measured for the first time in 2007 by the Parkes Telescope \cite{Lorimer2007}. Many models have been proposed to explain the burst's origin, but the emission mechanism remains unknown \cite{Platts2019}. The FRB's dispersion measure ($\mathrm{DM}$) is related to the density of free electrons along the line of sight from the source to the observer, and the large observed $\mathrm{DM}$, with respect to the Milk Way contribution, indicates an extragalactic or cosmological origin for the FRBs \cite{Dolag2015}. Although only a few FRBs in the literature are well localized (with redshift), $\mathrm{DM}$ can be used as an astrophysical and cosmological probe when combined with the redshift of the host galaxy of the FRB (see \cite{Walters2018,Wei2018,Lin2021,Wu2021,Reischke2023,Lemos2023,Lemos2023_2} for applications of FRBs in the cosmology). An important issue in FRB analyses is the poor knowledge about the dispersion measure's density fluctuations ($\delta$) due to the spatial variation in cosmic electron distribution. To circumvent this problem, such fluctuations can be parameterized as a function of redshift \cite{Takahashi2021}. Another issue when studying FRBs in cosmology is the host galaxy contribution of the FRBs ($\mathrm{DM}_{\mathrm{host}}$), which depends on factors such as the galaxy type, the mass of the host galaxy as well as the relative orientation between the FRB source and the host \cite{Xu2015}.

In a previous work \cite{Lemos2023}, we proposed and applied a cosmological model-independent method using FRB observations to constrain the baryon mass fraction in the intergalactic medium ($f_{\mathrm{IGM}}$). 
Here, we adapt and apply this method to test a possible time evolution of the fine-structure constant combining FRBs and SNe Ia observations. Using Monte Carlo simulations to investigate the potential of the method proposed when applied to larger samples of FRB measurements, we also forecast constraints on the $\Delta \alpha/\alpha$ relation of the runaway dilation scenario considered (Eq. \ref{eq:RunModel}). We organize this paper as follows. In Section \ref{sec:UsualDM}, we introduce the physical quantities related to FRBs, and derive the expressions for $\mathrm{DM}$ as a function of $\alpha$. In Section \ref{sec:DMcomp}, we present the composition of the dispersion measure and the cosmological-independent proposed method to test $\alpha$. Section \ref{sec:data} presents the data sets used in our analysis. The results obtained from the current FRB and SNe data are discussed in Section \ref{sec:results} while Section \ref{sec:simulations} describes the Monte Carlo simulations and our forecasts. We end the paper summarizing the main conclusions in Section \ref{sec:conclusions}.

%%%%%%%%%%%%%%%%%%%%%%%%%%%%%%%%%%%%%%%%%%%%%%%%%%%%%%%%%%%%%%%%%%%%%%%%%%%%%%%

\section{Theoretical Framework for FRBs}
\label{sec:UsualDM}

\subsection{Definition of Dispersion Measure}

In this section, we aim to search for a time variation of $\alpha$ based on the $\mathrm{DM}$ features. In what follows, we will review the well-established concepts of the FRBs' time delay and dispersion measure (see e.g. \cite{Rybick}). 

With effect, the time required for a pulse to reach the Earth from a source at a distance $d$ with frequency $\omega$ and group velocity $v_{g}$ is
\begin{equation}
\label{PlasmaTime}
t_{p} = \int_{0}^{d} \frac{dS}{v_{g}} \approx \int_{0}^{d} \left( 1 + \frac{\omega^{2}_{p}}{2\omega^{2}} \right) \frac{dS}{c} = \frac{d}{c} +  \int_{0}^{d} \frac{\omega^{2}_{p}}{2\omega^{2}c} dS \;\;
\end{equation}
where $S$ measures the line-of-sight distance from the source to the Earth and $\omega_p$ is the plasma frequency.

Considering that the frequency $\omega$ does not change along with $S$ and that the plasma frequency is given by
\begin{equation}
\label{eq:wp}
    \omega_{p}^{2} = \frac{4\pi e^{2} n_e}{m_e},
\end{equation}
where $e$ is the elementary charge, $m$ the electron mass, and $n_e$ is the electronic density of the plasma, Eq. \ref{PlasmaTime} gives
\begin{equation}
\label{TimePlasma}
t_{p} = \frac{d}{c} + \frac{2\pi e^{2}}{cm_e\omega^{2}} \int_{0}^{d} n_e dS.
\end{equation}
or still 
%The usually quantity measured is the rate of change of the arrival time with respect to frequency. So, the derivative of Eq. \ref{TimePlasma} gives us
%Replacing the plasma frequency relation (Eq. \ref{eq:wp}) into the above equation, and by assuming that the number density is the only quantity that changes with $S$, we find
\begin{equation}
\label{ArTime}
    \frac{dt_{p}}{d\omega} = - \frac{4\pi e^{2}}{cm_e\omega^{3}} \int_{0}^{d} n_e dS,
\end{equation}
which is the rate of change of the arrival time with respect to frequency. The dispersion measure ($\mathrm{DM}$) is defined as
\begin{equation}
    \mathrm{DM} = \int_{0}^{d} n_e dS.
\end{equation}

\subsection{Cosmological Extension}

From a cosmological point of view, it is necessary to redo the previous analysis by assuming that FRBs have an extragalactic origin at a given redshift $z$. Three effects must be taken into account \cite{Zhang2014}: the change in the observed frequency from the redshift of light ($\omega \rightarrow \omega = (1+z)\omega_{obs}$); a redshift dependence on the electronic density ($n_e = n_e(S) \rightarrow n_e = n_e(z)$); and the time dilation effect. Thus, we can rewrite Eq. (\ref{PlasmaTime}) as
\begin{equation}
\label{PlasmaTimeCosmo}
t_{p} = \frac{d}{c} + \frac{1}{2\omega_{\mathrm{obs}}^{2}} \int_{0}^{z} \frac{\omega^{2}_{p}}{(1+z)^2H(z')}dz' \; \; ,
\end{equation}
whose derivative with respect to the observed frequency is given by 
\begin{equation}
\label{eq:dt/dw}
    \frac{dt_{p}}{d\omega_{\mathrm{obs}}} = - \frac{1}{\omega_{\mathrm{obs}}^{3}} \int_{0}^{z}  \frac{\omega^{2}_{p}}{(1+z)^2H(z')}dz'\;,
\end{equation}
where $H(z)$ is the Hubble parameter. Note that we must assume a functional form for the plasma frequency from now on. Furthermore, it is important to observe that possible variations of physical parameters will arise from that term. As usual in the literature, the main focus is the electronic density, and the plasma frequency is assumed as Eq. (\ref{eq:wp}), thus
\begin{equation}\label{eq:dt/dwSt}
    \frac{dt_{p}}{d\omega_{\mathrm{obs}}} = - \frac{4\pi e^{2}}{cm_e\omega_{\mathrm{obs}}^{3}} \int_{0}^{z} \frac{n_e(z')c}{(1+z')^{2}H(z')} dz' \;,
\end{equation}
with the dispersion measure given by
\begin{equation}
    \mathrm{DM} = \int_{0}^{z}   \frac{n_e(z')c}{(1+z')^{2} H(z')}dz' \;.
\end{equation}

\subsection{Dispersion Measure as a function of $\alpha$}

The plasma frequency as a function of the fine-structure constant $\alpha$ can be written as
\begin{equation}
    \omega^{2}_{p} = \frac{2n_e h \alpha c}{m_e}\;.
\end{equation}
%then we replace it in Eq.~\ref{eq:dt/dw} in order to make the study dependent on the fine-structure constant. So, the change of the arrival time becomes
%\begin{equation}
%\label{eq:dt/dwAlfa}
%    \frac{dt_{p}}{d\omega_{obs}} = - \frac{1}{\omega^{3}_{obs}} \int_{0}^{z}  \frac{2 n_e h \alpha c}{m_e(1+z)^2H(z')}dz'.
%\end{equation}
Now, assuming that $n_e$ and $\alpha$ vary with respect to the redshift, Eq. (\ref{eq:dt/dwSt}) becomes
%\begin{equation}
%\label{eq:dt/dw2}
%    \frac{dt_{p}}{d\omega_{obs}} = - \frac{2h}{\omega^{3}_{obs} m_e} \int_{0}^{z}  \frac{n_e(z') \alpha (z') c}{(1+z)^2H(z')}dz'\;,
%\end{equation}
%or still 
%As we are looking for a definition of DM with a dependence on the fine-structure constant, we must compare the integrals in equations \ref{eq:dt/dw} and \ref{eq:dt/dwSt}, then we can notice that the difference is the $\alpha (z')$ factor. So, in order to obtain a mathematical structure from Eq. \ref{eq:dt/dwSt} that leads to \ref{eq:dt/dw} in the case where the fine-structure is constant, we can define the current fine-structure constant as $\alpha_{0} = \alpha (z=0)$, then one can multiplies Eq. \ref{eq:dt/dwSt} by a factor of $\alpha_{0}/\alpha_{0}$ which leads to
\begin{equation}
\label{eq:dt/dwComp}
    \frac{dt_{p}}{d\omega_{\mathrm{obs}}} = - \frac{2h \alpha_{0}}{\omega_{\mathrm{obs}}^{3} m_{e}} \int_{0}^{z}  \frac{n_e(z') c}{(1+z)^2H(z')} \frac{\alpha(z')}{\alpha_{0}} dz',
\end{equation}
where $\alpha_{0} = \alpha (z=0)$. Consequently, the dispersion measure considering a possible variation of the fine structure constant is given by
%\begin{equation}
%\label{eq:dt/dwAlpha}
%    DM(\alpha(z)) = \int_{0}^{z}  \frac{n_e(z') c}{(1+z)^2H(z')} \frac{\alpha(z')}{\alpha_{0}} dz' \; .
%\end{equation}
%Furthermore, as the parameterisations and measurements of variations of the fine structure constant are usually given in terms of the difference $\frac{\Delta \alpha(z)}{\alpha_0}$ then it is useful to rewrite the previous equation as 
\begin{equation}
\label{eq:dt/dwDelta}
    \mathrm{DM} = \int_{0}^{z}  \frac{n_e(z') c}{(1+z)^2H(z')} \left( \frac{\Delta \alpha(z')}{\alpha_{0}} + 1 \right) dz',
\end{equation}
where ${\Delta \alpha(z)} = \alpha(z) - \alpha_0$. As expected, the last two equations reduce to the standard case when $\alpha =$ const. In what follows, we will focus on a phenomenological approach, using FRBs observations, to constrain cosmological parameters.

%%%%%%%%%%%%%%%%%%%%%%%%%%%%%%%%%%%%%%%%%%%%%%%%%%%%%%%%%%%%%%%%%%%%%%%%%%%%%%%
%%%%%%%%%%%%%%%%%%%%%%%%%%%%%%%%%%%%%%%%%%%%%%%%%%%%%%%%%%%%%%%%%%%%%%%%%%%%%%%

\section{Dispersion Measure Components}
\label{sec:DMcomp}

Due to the interaction between photons from the burst and free electrons in the medium, the received radio waves experience a change in frequency, which in turn induces a change in their speed. Therefore, we can write the observed dispersion measure ($\mathrm{\mathrm{DM}}_{\mathrm{obs}}$) as a contribution of several components as \cite{Deng2014,Gao2014} 
\begin{equation}
\label{eq:DMobs}
\mathrm{DM}_{\mathrm{obs}}(z) = \mathrm{DM}_{\mathrm{ISM}} + \mathrm{DM}_{\mathrm{halo}} + \mathrm{DM}_{\mathrm{host}}(z) + \mathrm{DM}_{\mathrm{IGM}}(z),
\end{equation}
where the components are, respectively, the contributions from the Milky Way Interstellar Medium ($\mathrm{DM}_{\mathrm{ISM}} $), the Milky Way Halo ($\mathrm{DM}_{\mathrm{halo}}$), the host galaxy ($\mathrm{DM}_{\mathrm{host}}$) and the Intergalactic Medium ($\mathrm{DM}_{\mathrm{IGM}}$).

To perform a statistical comparison between the observational data and the theoretical parameters, we define the observed extragalactic dispersion measure 
\begin{equation}
\label{eq:DMext_obs}
  \mathrm{DM}^{\mathrm{obs}}_{\mathrm{ext}}(z) \equiv  \mathrm{DM}_{\mathrm{obs}}(z) - \mathrm{DM}_{\mathrm{ISM}}  - \mathrm{DM}_{\mathrm{halo}} \; ,
\end{equation}
and the theoretical extragalactic dispersion measure 
\begin{equation}
\label{DMext_th}
    \mathrm{DM}_{\mathrm{ext}}^{\mathrm{th}}(z) \equiv \mathrm{DM}_{\mathrm{host}}(z) + \mathrm{DM}_{\mathrm{IGM}}(z) \;.
\end{equation}
From the observational point of view, the $\mathrm{DM}_{\mathrm{obs}}(z)$ of a FRB is directly measured from the corresponding event. At the same time, $\mathrm{DM}_{\mathrm{ISM}}$ can be well-constrained using models of the ISM galactic electron distribution in the Milky Way from pulsar observations \cite{Taylor1993,Cordes2002,Yao2017}. Both values are observationally determined and can be found in Table \ref{tab:FRB}. %Moreover, the Milky Way halo contribution is not entirely understood yet, so we follow \cite{Macquart2020} and assume $DM_{halo} = 50$ pc/cm$^{3}$.% 
Moreover, the halo contribution of the Milky Way is not entirely understood yet, and we follow \cite{Macquart2020} and assume $\mathrm{DM}_{\mathrm{halo}} = 50$ pc/cm$^{3}$.

%%------------------------------------------------------  TABLE  ------------------------------------------------------

\begin{table*}
\centering
\caption{Properties of FRB with known host galaxies}
\begin{tabular}{ c  c  c  c  c c }
\hline
Name & Redshift $z$ & $\mathrm{DM}_{\mathrm{ISM}}$ & $\mathrm{DM}_{\mathrm{obs}}(z)$ & $\sigma_{\mathrm{obs}}$ & Reference\\
 & & [pc/cm$^{3}$] & [pc/cm$^{3}$] & [pc/cm$^{3}$] & \\
\hline
FRB 180916B & 0.0337  & 200.0 & 348.8  & 0.2 &\cite{FRB180916}\\
FRB 201124A & 0.098   & 123.2 & 413.52 & 0.5 &\cite{FRB201124}\\
FRB 190608B & 0.1178  & 37.2  & 338.7  & 0.5 &\cite{FRB190608}\\
FRB 200430A & 0.16    & 27.0    & 380.25 & 0.4 &\cite{FRB190523_2}\\
FRB 121102A & 0.19273 & 188.0 & 557.0  & 2.0 &\cite{FRB121102}\\
FRB 191001A & 0.234   & 44.7  & 506.92 & 0.04 &\cite{FRB190523_2}\\
FRB 190714A & 0.2365  & 38.0    & 504.13 & 2.0 &\cite{FRB190523_2}\\
FRB 191228A & 0.2432 & 33.0 & 297.5 & 0.05 & \cite{FRB20191228}\\
FRB 190102C & 0.291   & 57.3  & 363.6  & 0.3 &\cite{FRB190102}\\
FRB 180924B & 0.3214  & 40.5  & 361.42 & 0.06 &\cite{FRB180924}\\
FRB 180301A & 0.3305 &152.0 & 536.0 & 8.0 & \cite{FRB20191228}\\
FRB 200906A & 0.3688 & 36.0 & 577.8 & 0.02 & \cite{FRB20191228}\\
FRB 190611B & 0.378   & 57.83 & 321.4  & 0.2 &\cite{FRB190523_2}\\
FRB 181112A & 0.4755  & 102.0 & 589.27 & 0.03 &\cite{FRB181112}\\
FRB 190711A & 0.522   & 56.4  & 593.1  & 0.4 &\cite{FRB190523_2}\\
FRB 190523A & 0.66    & 37.0  & 760.8  & 0.6 &\cite{FRB190523_1,FRB190523_2}\\
FRB 210320 & 0.2797 & 42.2 & 384.8 & 0.3 & \cite{FRB210320}\\
%FRB 20220610A & 1.016 & 31.0 & 1458.15 & 0.2 &\cite{FRB20220610A} \\
\hline
%\multicolumn{6}{p{2.5cm}}{\,} 
\end{tabular}
\label{tab:FRB}
\end{table*}

Regarding the theoretical extragalactic dispersion measure, the host galaxy contribution is a poorly understood parameter due to the challenges in measurement and modeling. This difficulty arises particularly because it depends on the type of galaxy, the relative orientations of the FRBs source with respect to the host and source, and the near-source plasma \cite{Xu2015}. For this reason, we consider %follow \cite{Deng2014,Ioka2003} and we parameterize the redshift evolution of $DM_{host}(z)$ as $DM_{host,0}/(1+z)$, where the $(1+z)$ factor accounts for the cosmic dilation. We also introduce a function for possible variations of the fine-structure constant, i.e.
\begin{equation}
\label{DMhost}
\mathrm{DM}_{\mathrm{host}}(z) = \frac{\mathrm{DM}_{\mathrm{host},0}}{(1+z)}f(\alpha, z)\;,
\end{equation}
where the case $f(\alpha, z) = 1$ (hereafter named Fixed) corresponds to the usual approach that assumes contributions from all host galaxies are the same. 
%and assume two parameterizations for $DM_{host}(z)$. For the first case (named host 1), we follow  \cite{Lemos2023,Lemos2023_2} and consider that the contributions of each host galaxy are the same, which leads to $f(\alpha, z) = 1$. Besides this consideration may pose problems given the substantial variations observed among different FRBs, we prefer examining the influence of the $DM_{host}$ distribution in another study. This parameterization is written as
%\begin{equation}
%\label{eq:host1}
%DM_{host{ }1} (z) = \frac{DM_{host,0}}{(1+z)} \; .
%\end{equation}
We also consider a parameterization that takes into account the contribution from the fine-structure constant for each host galaxy. In this case (named $\alpha$-dependent host), considering the runaway dilaton model discussed earlier -- see e.g. Eqs. (\ref{eq:RunModel}) and (\ref{eq:dt/dwDelta}) --, we have $f(\alpha, z) = -\gamma \ln{(1+z)} + 1$.

%\begin{equation}
%\label{eq:host2}
%    DM_{host{ }2} (z) = \frac{DM_{host,0}}{(1+z)}\left(-\gamma \ln{(1+z)} + 1\right) \; .
%\end{equation}

The last term of the theoretical expression for the extragalactic dispersion measure is the average dispersion measure from the IGM (i.e. Eq. (\ref{DMext_th})), where cosmological contributions appear. Replacing in Eq. (\ref{eq:dt/dwDelta}) the electronic density $n(z)$, as given by \cite{Deng2014}, we obtain
%\begin{eqnarray}
%\label{DMigm}
%DM_{IGM}(z) &=& \frac{3c\Omega_{b}H_{0}^{2}}{8\pi Gm_{p}} \int_{0}^{z} \frac{(1+z')f_{IGM}(z')\chi(z')}{H(z')}  \nonumber\\
%&\times& \left( \frac{\Delta \alpha(z')}{\alpha_{0}} + 1 \right) dz'\;,
%\end{eqnarray}
\begin{equation}\label{DMigm}
\mathrm{DM}_{\mathrm{IGM}}(z) = \frac{3c\Omega_{b}H_{0}^{2}}{8\pi Gm_{p}} \int_{0}^{z} \frac{(1+z')f_{\mathrm{IGM}}(z')\chi(z')}{H(z')}  \left( \frac{\Delta \alpha(z')}{\alpha_{0}} + 1 \right) dz'\;,
\end{equation}
where $\Omega_{b}$ is the present-day baryon density parameter, $H_{0}$ is the Hubble constant, $m_{p}$ is the proton mass, $f_{\mathrm{IGM}}(z)$ is the baryon fraction in the IGM, and $\chi(z)$ is the free electron number fraction per baryon. This last quantity is given by

\begin{equation}
\chi(z) = Y_{H} \chi_{e,H}(z) + Y_{He} \chi_{e,He}(z)\;,
\end{equation}
where the terms $Y_{H} = 3/4$ and $Y_{He} = 1/4$ are the mass fractions of hydrogen and helium, respectively, while $\chi_{e,H}(z)$ and $\chi_{e,He}(z)$ are the ionization fractions of hydrogen and helium, respectively. At $z < 3$ hydrogen and helium are fully ionized ($\chi_{e,H}(z) = \chi_{e,He}(z) = 1$) \cite{Becker2011}, so that we have $\chi(z) = 7/8$.

A cosmological model is typically assumed to solve the integral for $\mathrm{DM}_{\mathrm{IGM}}(z)$. In our analysis, we follow the approach discussed in \cite{Lemos2023} and integrate Eq. (\ref{DMigm}) by parts using the definition of luminosity distance $d_{L}$. {This method allows us to avoid relying on any specific cosmological model, instead using an observed quantity: the supernovae luminosity distance.} Thus, by assuming the runaway dilaton model's parameterization for $\alpha$ and considering a constant $f_{\mathrm{IGM}}$ ($f_{\mathrm{IGM}} (z) = f_{\mathrm{IGM},0}$) \footnote{We are assuming that any intrinsic evolution of $f_{IGM}$, as well as $DM_{host}$, is not appreciable \cite{Theis2024}}., Eq. (\ref{DMigm}) can be written as
%\begin{eqnarray}\label{eq:IGM_alpha2}
%    DM_{IGM} (z) &=& A f_{IGM,0} \bigg[  \left(- \gamma \ln{(1+z)} + 1 \right) \frac{d_{L}(z)}{c}  \nonumber\\
%    &+& ( \gamma - 1)\int_{0}^{z} \frac{d_{L}(z')}{(1+z')c} dz'  \nonumber\\
%    &+& \gamma\int_{0}^{z} \frac{d_{L}(z')}{(1+z')c} \ln{(1+z')} dz' \bigg] \;, 
%\end{eqnarray}
\begin{eqnarray}\label{eq:IGM_alpha2}
    \mathrm{DM}_{\mathrm{IGM}} (z) &=& A f_{\mathrm{IGM},0} \bigg[  \left(- \gamma \ln{(1+z)} + 1 \right) \frac{d_{L}(z)}{c} + ( \gamma - 1)\int_{0}^{z} \frac{d_{L}(z')}{(1+z')c} dz'  \nonumber\\
   &+& \gamma\int_{0}^{z} \frac{d_{L}(z')}{(1+z')c} \ln{(1+z')} dz' \bigg] \;, 
\end{eqnarray}
where $A= \frac{21c\Omega_{b}H^{2}_{0}}{64\pi Gm_{p}}$. The above integrals can be numerically solved as (see \cite{Holanda2013}):
%\int_{0}^{z} \frac{d_{L}(z')}{(1+z')c} dz' &=&  \frac{1}{2c}\sum_{i=1}^{N} \left( z_{i+1}-z_{i}\right)  \nonumber\\&\times& \left[\frac{d_{L}(z_{i+1})}{(1+z_{i+1})} 
%+ \frac{d_{L}(z_{i})}{(1+z_{i})}  \right]\;,
%\end{eqnarray}
\begin{equation}\label{soma}
\int_{0}^{z} \frac{d_{L}(z')}{(1+z')c} dz' = \frac{1}{2c}\sum_{i=1}^{N} \left( z_{i+1}-z_{i}\right)  \times \left[\frac{d_{L}(z_{i+1})}{(1+z_{i+1})} 
+ \frac{d_{L}(z_{i})}{(1+z_{i})}  \right]\;,
\end{equation}
and
%\begin{eqnarray}\label{soma}
%\int_{0}^{z} \frac{d_{L}(z')}{(1+z')c} \ln{(1+z')} dz' &=&  \frac{1}{2c}\sum_{i=1}^{N} \left( z_{i+1}-z_{i}\right)\nonumber\\ &\times& \bigg[ \frac{d_{L}(z_{i+1})}{(1+z_{i+1})} \ln{(1+z_{i+1})} \nonumber \\
%&+& \frac{d_{L}(z_{i})}{(1+z_{i})} \ln{(1+z_{i})}  \bigg]\;.
%\end{eqnarray}

\begin{eqnarray}\label{somaII}
\int_{0}^{z} \frac{d_{L}(z')}{(1+z')c} \ln{(1+z')} dz' &=&  \frac{1}{2c}\sum_{i=1}^{N} \left( z_{i+1}-z_{i}\right) \times \bigg[ \frac{d_{L}(z_{i+1})}{(1+z_{i+1})} \ln{(1+z_{i+1})} \nonumber \\
&+& \frac{d_{L}(z_{i})}{(1+z_{i})} \ln{(1+z_{i})}  \bigg]\;.
\end{eqnarray}

From the above expressions, one can constrain a possible evolution of the baryon fraction by modeling both $\mathrm{DM}_{\mathrm{host},0}$ and $\mathrm{DM}_{\mathrm{IGM}}$ and comparing the theoretical predictions with the observed values of $\mathrm{DM}_{\mathrm{ext}}$. Specifically, by combining FRBs with SNe dataset we can obtain cosmological model-independent constraints on $\gamma$, $f_{\mathrm{IGM},0}$ and $\mathrm{DM}_{\mathrm{host},0}$, from the observational data sets described below.   

%%%%%%%%%%%%%%%%%%%%%%%%%%%%%%%%%%%%%%%%%%%%%%%%%%%%%%%%%%%%%%%%%%%%%%%%%%%%%%%%%%%%%%%%%%%%%%%%%%%%%%%%%%%%%%%%%

\section{Data sets}
\label{sec:data}

\subsection{FRBs}

The most up-to-date FRB data set currently available comprises 23 well-localized events of FRBs (see \cite{2021ascl.soft06028S} for details of FRBs catalog\footnote{https://www.herta-experiment.org/frbstats/}). In our analysis, we exclude the following events: FRB 190614 \cite{FRB190614}, as it has no measurement of spectroscopic redshift and can be associated with two host galaxies; FRB 200110E \cite{Bhardwaj2021} is estimated to be in the direction of M81, but a Milky Way halo origin can not be rejected; FRB 181030 \cite{Bhardwaj2021_2} has a low-redshift ($z = 0.0039$) and it can not be associated with any SNe in the Pantheon catalog; FRB 20190520B \cite{Ocker} has a host galaxy contribution significantly larger than the other events; FRB 210117 \cite{FRB210320} has a observed $\mathrm{DM}$ much larger for its redshift ($z = 0.2145$); and finally, FRB 20220610A \cite{FRB20220610A} at $z = 1.016$, which will be removed from our sample for reasons discussed in Sec. \ref{sec:simulations}.%\ref{sec:FRBz=1.016}. 

We list in Table \ref{tab:FRB} our working sample that contains 17 FRBs \cite{FRB210320,FRB180916,FRB201124,FRB190608,FRB190523_2,FRB121102,FRB20191228,FRB190102,FRB180924,FRB181112,FRB190523_1} and their main properties: redshift, the Galaxy contribution ($\mathrm{DM}_{\mathrm{MW,ISM}}$) estimated from the NE2001 model \cite{Cordes2002}, observed dispersion measure ($\mathrm{DM}_{\mathrm{obs}}$), $\mathrm{DM}_{\mathrm{obs}}$ uncertainty ($\sigma_{\mathrm{obs}}$) and the references. 

%The observational quantity $DM_{\mathrm{ext}}$ can be obtained from table \ref{tab:FRB} and its uncertainty can be calculated by the expression 

The observational quantity $\mathrm{DM}_{\mathrm{ext}}$ can be obtained from Table \ref{tab:FRB} and its total uncertainty can be expressed by the relation
\begin{equation}\label{uncertainty}
    \sigma_{\mathrm{tot}}^{2} = \sigma_{\mathrm{obs}}^{2} + \sigma_{\mathrm{MW}}^{2} + \sigma_{\mathrm{IGM}}^{2} + \bigg( \frac{\sigma_{\mathrm{host},0}}{1+z} \bigg)^{2} + \delta^{2} \;,
\end{equation}
where $\sigma_{\mathrm{obs}}$ can be obtained from Table \ref{tab:FRB}, the average galactic uncertainty $\sigma_{\mathrm{MW}}$ is assumed to be 10 pc/cm$^{3}$ \cite{Manchester2005} and $\delta$ is the fluctuations of $\mathrm{DM}$ which is related to the spatial variation in cosmic electron density along the line-of-sight. Such fluctuations are not currently well determined by observations. Therefore, we will treat them as a fixed value, $\delta = 230\sqrt{z}$ pc/cm$^{3}$ \cite{Takahashi2021,Lemos2023}, in the statistical analyses. Following reference \cite{Li2019}, we adopted $\sigma_{\mathrm{host},0} = 30$ pc/cm$^{3}$ as the uncertainty of $\mathrm{DM}_{\mathrm{host},0}$. Furthermore, the uncertainty of IGM contribution ($\sigma_{\mathrm{IGM}}$) can be calculated from error propagation of Eq. \ref{eq:IGM_alpha2} given by

\begin{eqnarray}
    \sigma_{\mathrm{IGM}} = A f_{\mathrm{IGM},0} \bigg[  \left(- \gamma \ln{(1+z)} + 1 \right)^{2} \frac{\sigma_{d_{L}}^{2}}{c^{2}} + ( \gamma - 1)^{2} \sigma_{\mathrm{I}}^{2} + \gamma^{2} \sigma_{\mathrm{II}}^{2}  \bigg]^{1/2},
\end{eqnarray}
where $\sigma_{d_{L}}$ is the luminosity distance uncertainty that is calculated from the SNe, $\sigma_{\mathrm{I}}$ and $\sigma_{\mathrm{II}}$ are the uncertainty of the quantities in Eqs. \ref{soma} and \ref{somaII}, respectively.

\subsection{Type IA Supernovae}

We use SNe observations from Pantheon catalog \cite{Scolnic} which comprises 1048 SNe within the redshift range $0.01 < z < 2.3$. We obtain $d_{L}$ from the distance moduli ($\mu(z)$) relation
\begin{equation} \label{mz}
    \mu(z) = m_{B} - M_{B} = 5\log_{10}\left[ \frac{d_{L}(z)}{1\mbox{Mpc}}\right] + 25 \;,
\end{equation}
where $m_{B}$ and $M_{B}$ are the apparent and absolute magnitude, respectively. In our analysis we fix $M_{B} = -19.214 \pm 0.037$ mag \cite{Riess2019}. The luminosity distance uncertainty can be expressed by relation

\begin{equation}
    \sigma_{d_{L}} = \frac{\ln{10}}{5}d_{L} \cdot \sqrt{\sigma_{m_{B}}^{2} + \sigma_{M_{B}}^{2}},
\end{equation}
being $\sigma_{m_{B}}$ and $\sigma_{M_{B}}$ the apparent and absolute magnitude uncertainties, respectively.

%\begin{equation}
% \sigma_{\mu} = \sqrt{\sigma_{m_{B}}^{2} + \sigma_{M_{B}}^{2}}. 
%\end{equation}

To estimate $d_{L}$ and its uncertainty at the same redshift of FRBs, we follow \cite{Lemos2023} and perform a Gaussian Process (GP) reconstruction of the Pantheon data, using GaPP python library\footnote{For details of GaPP (https://github.com/astrobengaly/GaPP), see \cite{GaPP}.}.

\begin{figure*}
\begin{center}
\includegraphics[width=0.44\textwidth]{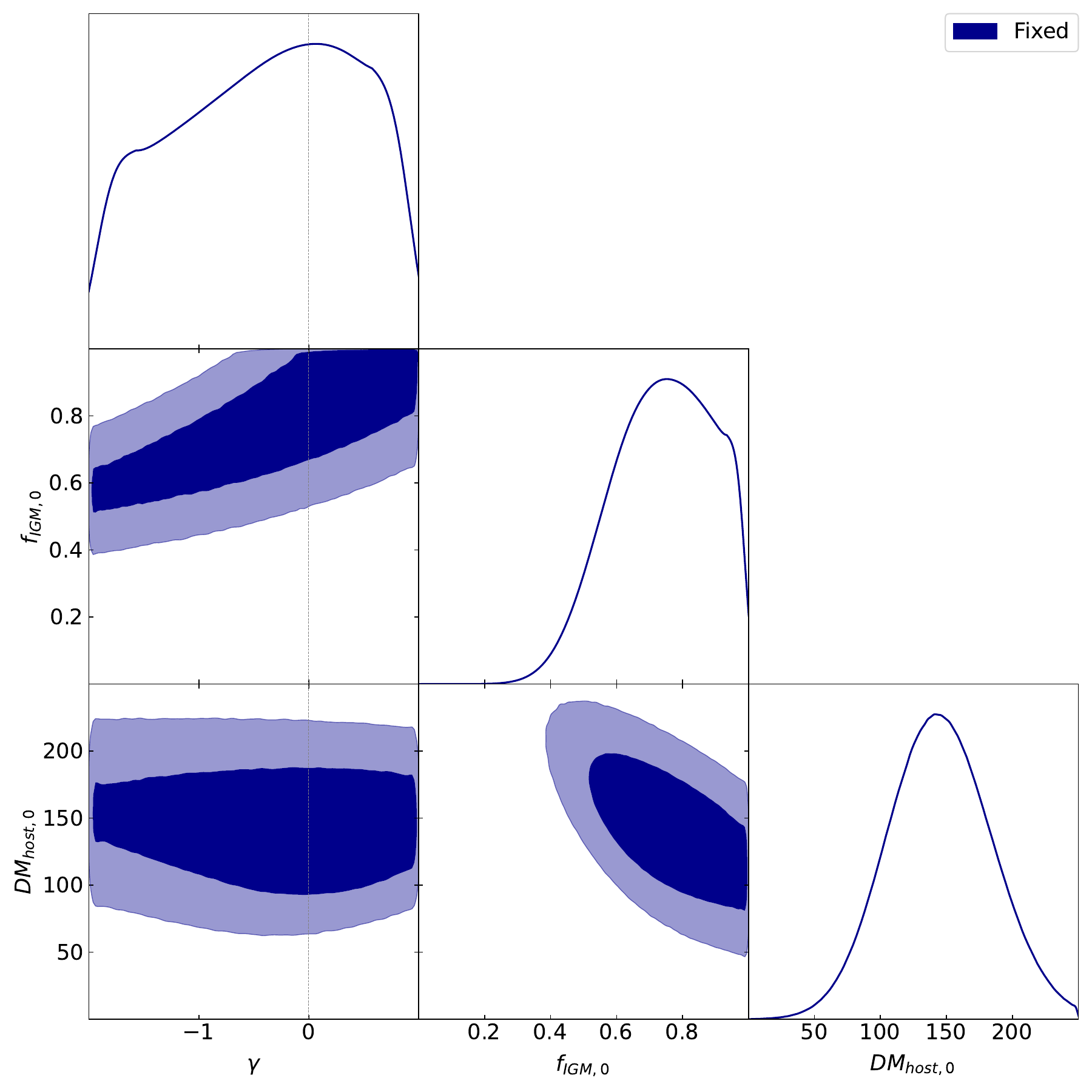}
\includegraphics[width=0.44\textwidth]{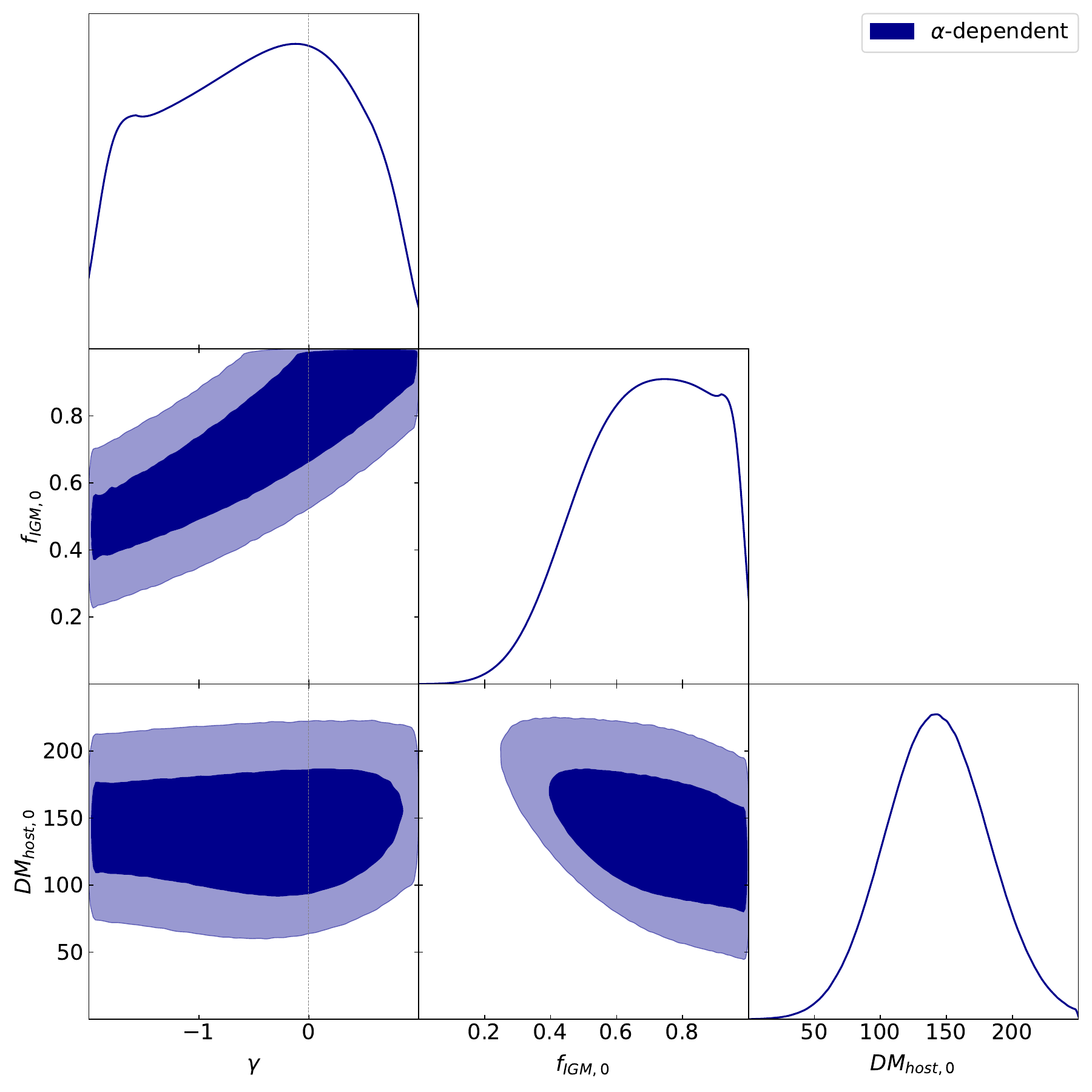}
%\vspace{-0.5cm}
\caption{Constraints on the parameter of runaway dilaton model $\gamma$, the baryon fraction $f_{\mathrm{IGM}}$ and the host galaxy contribution $\mathrm{DM}_{\mathrm{host},0}$ for parameterizations {{\textbf{Fixed host}}} ({\it{left}}) and {{\textbf{$\alpha$-dependent host}}} ({\it{right}}).} 
\label{fig:results_obs}
%colour_magnitude_diagram
\end{center} 
\end{figure*}

\section{Results}
\label{sec:results}

We perform a Monte Carlo Markov Chain (MCMC) analysis using the \textit{emcee} package \cite{Foreman-Mackey2013} to constrain the runaway dilaton model's parameter $\gamma$, $f_{\mathrm{IGM},0}$ and $\mathrm{DM}_{\mathrm{host},0}$. In the analysis, we assume $\Omega_{b}h^{2} =  0.02235 \pm 0.00037$, as reported in \cite{Cooke2018}. %The results of our observational data analysis are displayed in Table \ref{tab:results_obs}.

Figure \ref{fig:results_obs} shows the posterior probability density function and $1 - 2 \sigma$ contours for combinations of the parameters $\gamma$, $f_{\mathrm{IGM},0}$ and $\mathrm{DM}_{\mathrm{host},0}$, considering parametrizations Fixed host (left Panel) and $\alpha$-dependent host (right Panel). The numerical results are presented in Table \ref{tab:results_obs}. As physically expected, both parameterizations have little impact on the estimates of $f_{\mathrm{IGM},0}$ and $\mathrm{DM}_{\mathrm{host},0}$, and show results consistent with $\Delta \alpha/\alpha =0$. However, in comparison with other cosmological constraints on the time-evolution of $\alpha$, which place constraints on $\Delta \alpha/\alpha$ ranging from $10^{-2}$ to $10^{-7}$ over a large redshift interval (see, e.g. \cite{Albareti_2015,Kotuš_2016,Holanda:2017qya,Marsh_2017,Holanda:2015oda,Lopez_2020,Zhang_2018,Colaco:2019fvl,Landau_2019,Rodrigo_2020}), the results shown in Table \ref{tab:results_obs} reflect the limitation of the current FRB observations in tightly constraining cosmological parameters. In what follows, we simulate and forecast the constraining power of larger samples of FRBs to limit a possible time variation of $\alpha$.

\begin{table}
%\begin{sidewaystable}
\centering
%\begin{center}
%\scalebox{0.83}{
\renewcommand{\arraystretch}{1.3}
\begin{tabular}{|c|c|c|c|}
%\hline
\hline
 & $\gamma$ & $f_{\mathrm{IGM},0}$ & $\mathrm{DM}_{\mathrm{host},0}$   \\
 &          &             & [pc/cm$^{3}$]  \\
\hline
\hline

Fixed host & $+0.04^{+0.47}_{-1.40}$ & $0.75^{+0.16}_{-0.16}$ & $142^{+42}_{-36}$\\

$\alpha$-dependent host & $-0.07^{+0.44}_{-1.39}$ & $0.75^{+0.15}_{-0.25}$ & $142^{+40}_{-37}$ \\

\hline
\end{tabular}
%}
\caption{Results for $\gamma$, $f_{\mathrm{IGM},0}$ and $\mathrm{DM}_{\mathrm{host},0}$ using the current FRB and SNe data and considering two cases of host contribution. The error bars correspond to $1\sigma$ level.}
\label{tab:results_obs}
%\end{center}
%\end{sidewaystable} 
\end{table}

%%%%%%%%%%%%%%%%%%%%%%%%%%%%%%%%%%%%%%%%%%%%%%%%%%%%%%%%%%%%%%%%%%%%%%%%%%%%%%%%%%%%%%%%%%%%%%%%%%%%%%%%%%%%%%%%%

\section{Simulations}
\label{sec:simulations}

As mentioned earlier, it is difficult to identify the host galaxy of the burst, so there are only a few FRBs in the literature with measured redshift. In order to study the impact of a larger sample of FRBs in constraining $\alpha$, we perform a Monte Carlo simulation to generate random points of $\mathrm{DM}_{\mathrm{ext}}$ from two different redshift distribution models (see \cite{Lemos2023_2} for more explanation about the impact of the redshift distribution models of FRBs). We will consider the distributions below:

\begin{itemize}
    
    \item {\bf{Gamma-Ray Bursts:}} Several studies assume the gamma-ray burst distribution for FRBs due to the similarities between these two events \cite{GRB}. In this case, the density function is given as %(termed 'PGRB'). 
\begin{equation}
    P_{\rm{GRB}}(z) \propto z  e^{-z}.
\end{equation}

    \item {\bf{Star Formation Rate:}} The star formation rate distribution was proposed by \cite{Madau} (for the first proposal of redshift distribution for FRBs, see reference \cite{Bhattacharya}). The spatial distribution of FRBs is expected to closely trace the cosmic one for young stellar FRB progenitors. The cosmic SFR function is written as
\begin{equation}
    \psi(z) = 0.015 \frac{(1+z)^{2.7}}{1+[(1+z)/2.9]^{5.6}} . 
\end{equation}

\end{itemize}

\begin{table*}
%\begin{sidewaystable}
\centering
%\begin{center}
\scalebox{0.89}{
\renewcommand{\arraystretch}{1.3}
\begin{tabular}{|c|c|c|c|c|c|c|}
%\hline
\hline
 & $\gamma$ & $f_{\mathrm{IGM},0}$ & $\mathrm{DM}_{\mathrm{host},0}$  & $\gamma$ & $f_{\mathrm{IGM},0}$ & $\mathrm{DM}_{\mathrm{host},0}$  \\
 &          &             & [pc/cm$^{3}$]  &          &             & [pc/cm$^{3}$]  \\
\hline
 & \multicolumn{3}{|c|}{SFR}               & \multicolumn{3}{|c|}{GRB} \\ 
%& SFR &  &  & GRB & Uniform & & ED &  \\ 
\hline
\hline
 &          &             &      $ N = 500 $          &          &             &   \\  
\hline

Fixed & $+0.0888^{+0.0365}_{-0.0972}$ & $0.9993^{+0.0062}_{-0.0570}$ & $125.20^{+21.31}_{-15.86}$ & $+0.2327^{+0.0477}_{-0.1997}$ & $0.9733^{+0.0189}_{-0.1012}$ & $110.21^{+21.32}_{-16.41}$ \\

$\alpha$-dependent & $+0.0197^{+0.0338}_{-0.1069}$ & $0.9991^{+0.0067}_{-0.0641}$ & $133.13^{+17.46}_{-15.99}$ & $+0.0138^{+0.0350}_{-0.1508}$ & $0.9990^{+0.0095}_{-0.0871}$ & $132.07^{+17.46}_{-15.55}$ \\

\hline

\hline
 &          &             &      $ N = 1000 $          &          &             &   \\  
\hline

Fixed & $+0.0856^{+0.0364}_{- 0.0938}$  & $0.9993^{+0.0058}_{-0.0548}$  & $123.20^{+14.90}_{-15.86}$  & $+0.2678^{+0.0318}_{- 0.1262}$  & $0.9868^{+0.0111}_{-0.0722}$  & $105.47^{+14.90}_{-11.49}$  \\

$\alpha$-dependent& $+ 0.0338^{+ 0.0240}_{-0.0571}$ & $0.9995^{+0.0036}_{-0.0351}$ & $131.35^{+11.73}_{-11.32}$ & $+0.0317^{+0.0261}_{-0.0785}$ & $0.9993^{+0.0050}_{-0.0478}$ & $128.38^{+11.73}_{-10.92}$ \\

\hline
\end{tabular}
}
\caption{The results of our simulations for $\gamma$, $f_{\mathrm{IGM},0}$ and $\mathrm{DM}_{\mathrm{host},0}$ considering the distribution models discussed in the text. }
\label{tab:results_sim}
%\end{center}
%\end{sidewaystable} 
\end{table*}

The steps of our simulations are the following (see also \cite{Lemos2023_2}): 
\begin{enumerate}
    \item  We generate random points using the redshift distribution models described above in the redshift range $[0.022, 1.5]$, considering FRB samples with $N = 500$ and $1000$ points. 

    \item The fiducial $\mathrm{DM}_{\mathrm{ext}}$ ($\mathrm{DM}_{\mathrm{ext}}^{\mathrm{fid}}$) is calculated using Eq. (\ref{DMext_th}), where $\mathrm{DM}_{\mathrm{IGM}}$ is given by Eq. (\ref{DMigm}). We adopt as fiducial values for the mean values of baryon fraction and host contribution the results we obtained above for the case $\gamma =0$. In our simulations, we also adopt the values of $H_{0} = 74.03 \pm 1.4$ kms$^{-1}$Mpc$^{-1}$ \cite{Riess2019}, $\Omega_{m} = 0.3153 $ \cite{Planck} and $\Omega_{b}h^{2} =  0.02235 \pm 0.00037$ \cite{Cooke2018}.

    \item We calculate the uncertainty of the simulated $\mathrm{DM}_{\mathrm{ext}}$ ($\sigma_{\mathrm{ext}}^{\mathrm{sim}}$), by performing a hyperbolic regression fit of the observational relative error ($\eta =\sigma_{\mathrm{ext}}^{\mathrm{obs}}/\mathrm{DM}_{\mathrm{ext}}^{\mathrm{obs}}$), 
    given by $\eta = A/z $, where $A$ is a hyperbolic regression free parameter.

    \item  The standard deviation of the Gaussian Distribution ($sd$) is obtained from the average distance between the observed and fiducial points.

    \item Finally, we calculate the simulated $\mathrm{DM}_{\mathrm{ext}}$ by assuming a normal distribution, given by $\mathrm{DM}_{\mathrm{ext}}^{\mathrm{sim}} (z) = \mathcal{N}(\mathrm{DM}_{\mathrm{ext}}^{\mathrm{fid}},sd)$. 
    
\end{enumerate}
We perform the steps above 100 times for each sample size of the distribution models, which is enough to obtain convergence (see Appendix A). In each simulation, we calculate the best fit of the free parameters and, subsequently, the average of each ensemble of 100 simulations. 

%%%%%%%%%%%%%%%%%%%%%%%%%%%%%%%%%%%%%%%%%%%%%%%%%%%%%%%%%%%%%%%%%%%%%%%%%%%%%%%%%%%%%%%%%%%%%%%%%%%%%%%%%%%%%%%%%
\begin{figure}
\begin{center}
\includegraphics[width=0.65\textwidth]{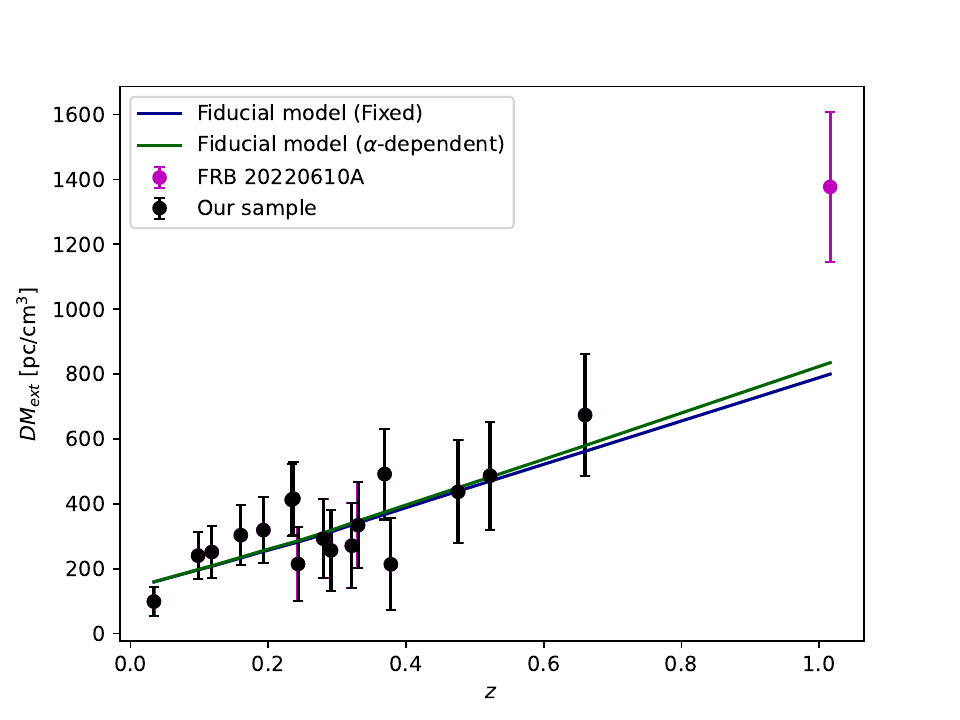}
\caption{The evolution of $\mathrm{DM}_{\mathrm{ext}}$ with redshift. Black points correspond to the 17 FRBs observations used in our analysis, while the pink point stands for the FRB 20220610A. Blue and green lines represent \textbf{Fixed host} and \textbf{$\alpha$-dependent host} parameterizations, respectively. As discussed in the text, the fiducial model is calculated using Eq. (\ref{DMext_th}), where $\mathrm{DM}_{\mathrm{IGM}}$ is given by Eq. (\ref{DMigm}). } 
\label{fig:ext18pts}
\end{center}
\end{figure}
%%%%%%%%%%%%%%%%%%%%%%%%%%%%%%%%%%%%%%%%%%%%%%%%%%%%%%%%%%%%%%%%%%%%%%%%%%%%%%%%%%%%%%%%%%%%%%%%%%%%%%%%%%%%%%%%%

Table \ref{tab:results_sim} shows the numerical values obtained separately for all distributions and different numbers of points in each realization ($N = 500$ and 1000) for both host cases. For the redshift distribution models considered, the constraints on $\gamma$ are significantly stronger than the current limits displayed in Table \ref{tab:results_obs}, with the error bar reaching $10^{-2}$ in most cases. It is also worth mentioning that the results do not show significant improvements on the limits to $\gamma$ (the same also happens to $f_{\mathrm{IGM},0}$ and $\mathrm{DM}_{\mathrm{host},0}$) from $N = 500$ to $N = 1000$ cases, which may indicate that to improve the constraints on $\Delta \alpha/\alpha$ further (beyond $\sigma \simeq 10^{-2}$), the quality of the FRB data will play a crucial role in the future analyses.  

Finally, as mentioned in Sec. \ref{sec:data}, we removed from our analysis the FRB 20220610A \cite{FRB20220610A} at $z = 1.016$. This FRB, observed in 2022 by the ASKAP (Australian Square Kilometre Array Pathfinder \cite{ASKAP}), has a very high dispersion measure (see Table \ref{tab:FRB}). We performed tests including this FRB in our analysis and found significantly different values of the parameter of the runaway dilaton model for Fixed host and $\alpha$-dependent host, i.e., $\gamma = -1.00^{+0.99}_{-0.15}$ and  $\gamma = -0.96^{+0.92}_{-0.12}$ at 1$\sigma$, respectively. As shown in Figure \ref{fig:ext18pts}, these results are inconsistent with the values found considering the sample with 17 FRBs (see Table \ref{tab:results_obs}), and that was the reason which led us to remove it from our analysis.

%%%%%%%%%%%%%%%%%%%%%%%%%%%%%%%%%%%%%%%%%%%%%%%%%%%%%%%%%%%%%%%%%%%%%%%%%%%%%%%%%%%%%%%%%%%%%%%%%%%%%%%%%%%%%%%%%%%%%
\section{Conclusions}
\label{sec:conclusions}

The question of a possible time dependence on the fundamental constants of Nature is a long-standing problem in theoretical physics. Concerning the fine-structure constant, a definitive answer may be provided with the entry into operation of very high-resolution spectrographs, such as ELT-ANDES  \cite{ANDES:2023cif}. Other types of cosmological observations can also play an important role in this discussion, given their ability to test variations of these constants in different redshift ranges.

This paper presented a cosmological model-independent test of a possible evolution of $\alpha$ with redshift from FRBs and SNe observations. We derived all the relevant expressions for the analysis and considered the runaway dilaton scenario, for which $\frac{\Delta \alpha}{\alpha} = - \gamma\ln{(1+z)}$. Without assuming any cosmological model, our analysis constrained the parameter $\gamma$ combining measurements of 17 well-localized FRBs and the Pantheon SNe compilation, following the method originally reported in \cite{Lemos2023}. As displayed in Table \ref{tab:results_obs}, we obtained results consistent with no variation of $\alpha$ and errors of the order of $10^{-1}$, reflecting the current data's limitation to impose tight limits on $\alpha(z)$.

Our analysis also included Monte Carlo simulations to forecast the potential of the method proposed when applied to larger samples of FRB measurements. We extended the samples to $N = 500$ and $N = 1000$ data points and analyzed two distribution models for the FRBs: Star Formation Rate (SFR) and Gamma Ray Burst (GRB). The results demonstrated that the uncertainties on $\gamma$ can be improved by one order of magnitude ($\sigma \sim 10^{-2}$), making it a competitive test when compared to other cosmological probes at the same redshift range. The lack of significant differences in the results regarding the number of points suggests that limits on $\frac{\Delta \alpha}{\alpha}$ beyond $\sigma \sim 10^{-2}$ will depend crucially on the quality of upcoming FRB data. This analysis underscores the potential of the proposed method to impose constraints on a possible redshift evolution of the fine-structure constant with larger and more precise samples of FRBs, as expected from current and planned observational projects (see e.g. \cite{SKA,ASKAP_CRACO,CHIME_outrigguers}).

%%-----------------------------------------------------------------------------

\section*{Acknowledgements}

TL thanks the financial support from the Coordena\c{c}\~ao de perfei\c{c}oamento de Pessoal de N\'{\i}vel Superior (CAPES). RSG thanks financial support from the Funda\c{c}\~ao de Amparo \`a Pesquisa do Estado do Rio de Janeiro (FAPERJ) grant SEI-260003/005977/2024 - APQ1. JSA is supported by Conselho Nacional de Desenvolvimento Cient\'{\i}fico e Tecnol\'ogico (CNPq No. 307683/2022-2) and Funda\c{c}\~ao de Amparo \`a Pesquisa do Estado do Rio de Janeiro (FAPERJ) grant 259610 (2021). This work was developed thanks to the High-Performance Computing Center at the National Observatory (CPDON).

%%-----------------------------------------------------------------------------
%\section*{Data Availability}

%The inclusion of a Data Availability Statement is a requirement for articles published in MNRAS. Data Availability Statements provide a standardised format for readers to understand the availability of data underlying the research results described in the article. The statement may refer to original data generated in the course of the study or to third-party data analysed in the article. The statement should describe and provide means of access, where possible, by linking to the data or providing the required accession numbers for the relevant databases or DOIs.

%%%%%%%%%%%%%%%%%%%% REFERENCES %%%%%%%%%%%%%%%%%%

%%%%%%%%%%%%%%%%%%%%%%%%%%%%%%%%%%%%%%%%%%%%%%%%%%

%%%%%%%%%%%%%%%%% APPENDICES %%%%%%%%%%%%%%%%%%%%%
\onecolumn
\appendix
\section{\centering Supplementary Material} \label{Appendix}

%\appendix{}
%\section{Supplementary Figures}\label{Appendix}

\begin{figure*}[!h]
\begin{center}
\resizebox{200pt}{130pt}{\includegraphics{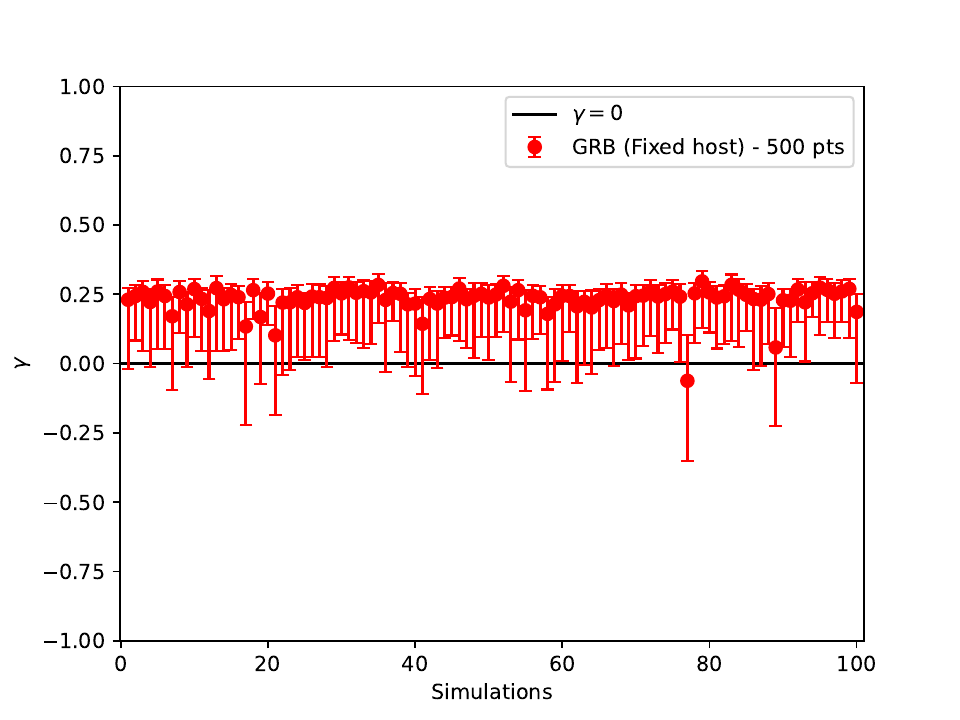}}
\hspace{1mm}\resizebox{200pt}{130pt}{\includegraphics{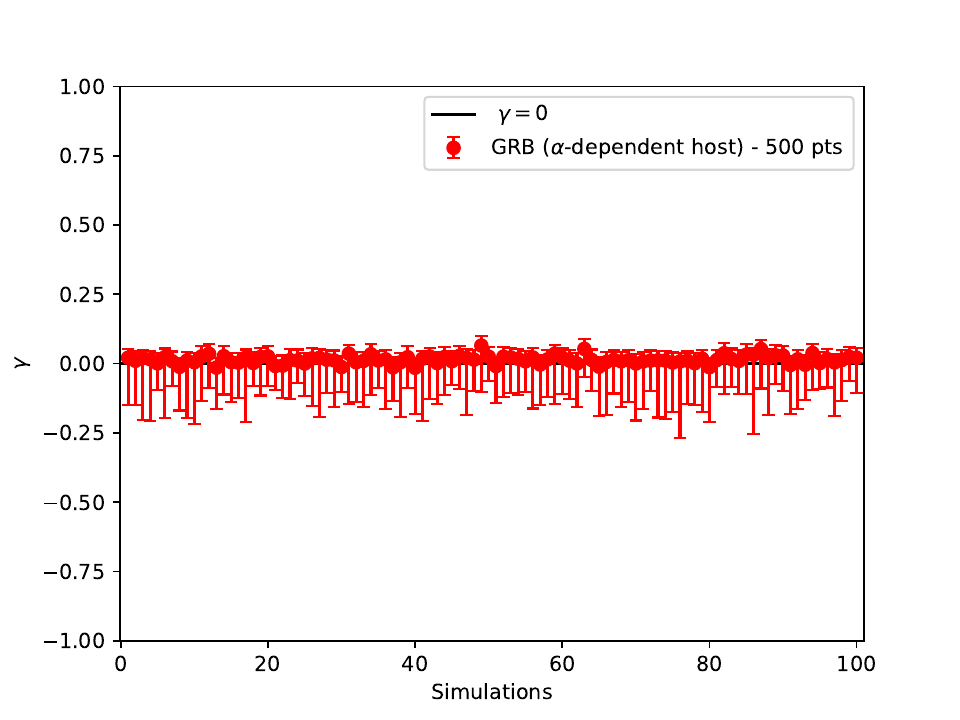}}\\
\hspace{1mm}\resizebox{200pt}{130pt}{\includegraphics{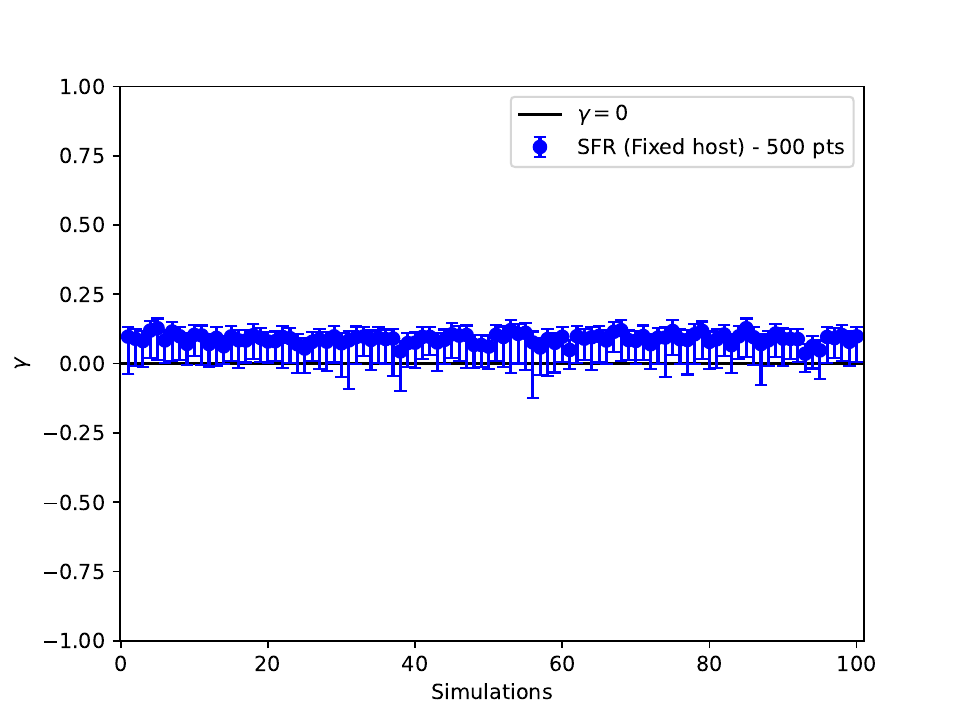}}
\resizebox{200pt}{130pt}{\includegraphics{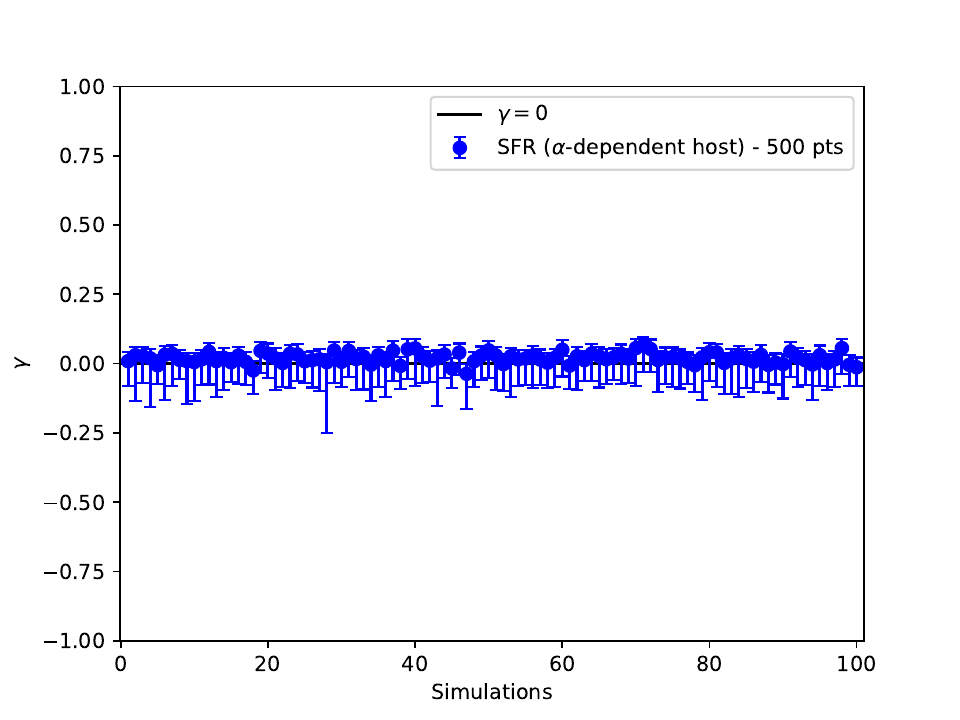}}\\
\end{center}
\caption{The best-fit of the 500 simulations of $\gamma$ for GRB and SFR distributions considering $N = 500$ for \textbf{Fixed host} and \textbf{$\alpha$-dependent host}. }  
\label{lowexp}
\end{figure*}

\begin{figure*}[!h]
\begin{center}
\resizebox{200pt}{130pt}{\includegraphics{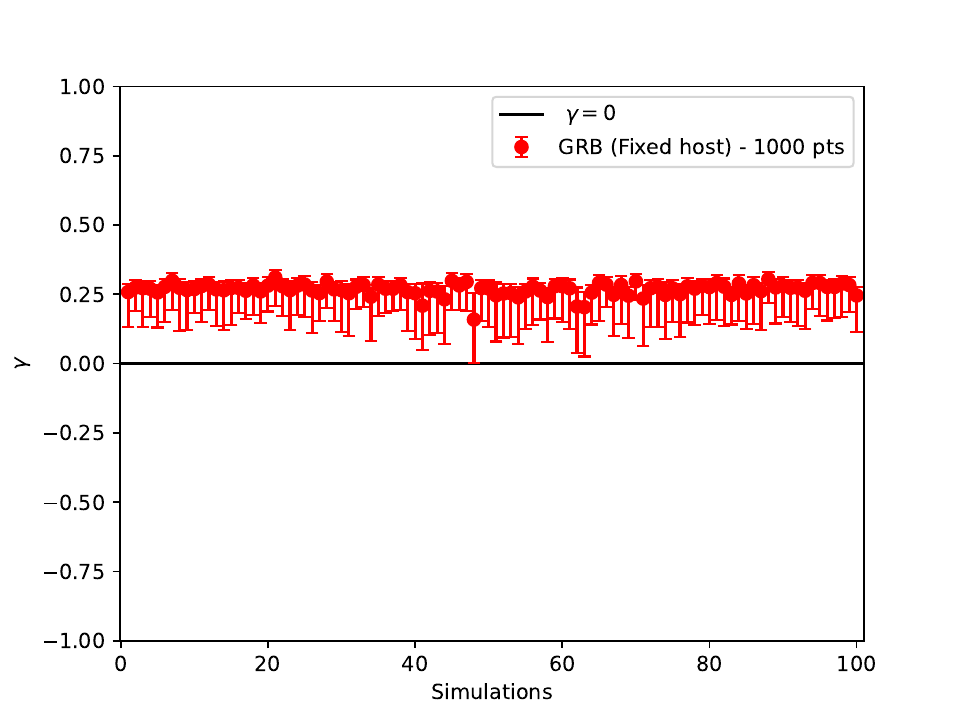}}
\hspace{1mm}\resizebox{200pt}{130pt}{\includegraphics{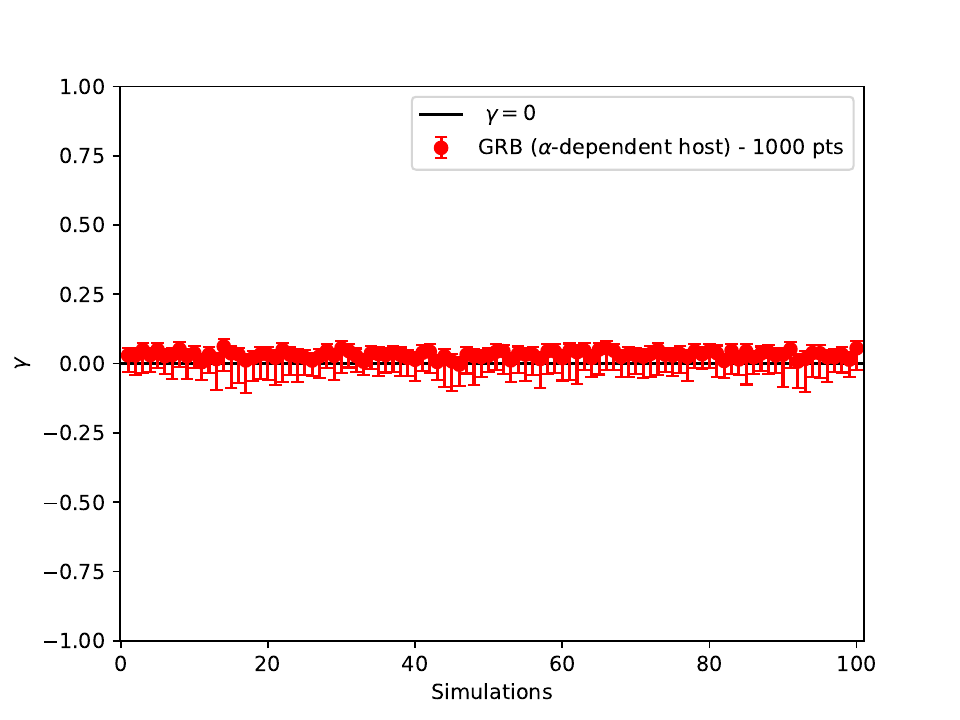}}\\
\hspace{1mm}\resizebox{200pt}{130pt}{\includegraphics{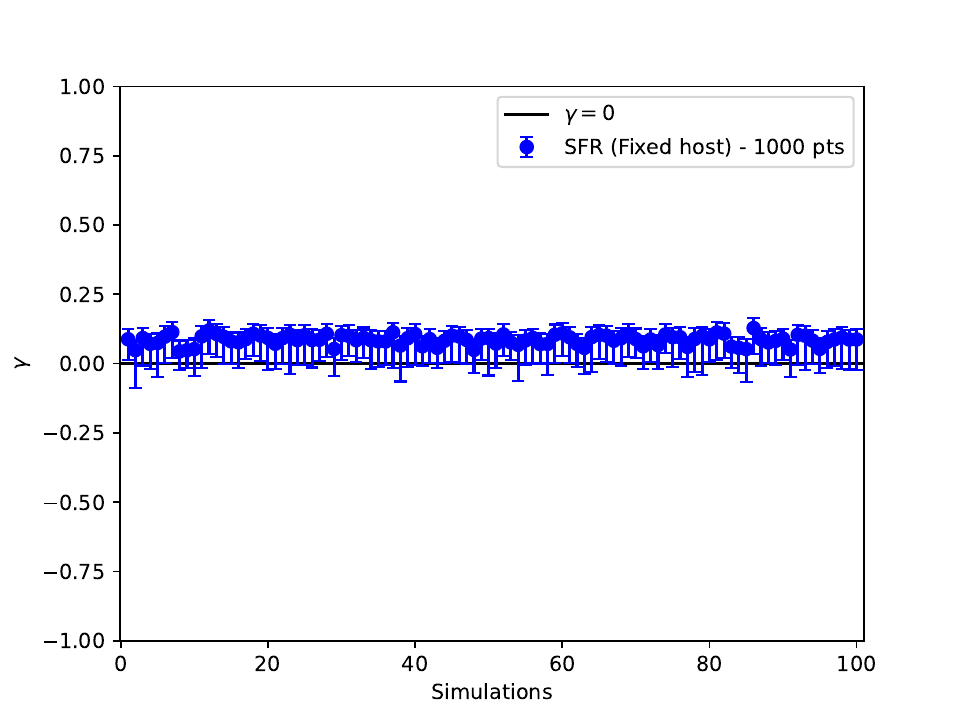}}
\resizebox{200pt}{130pt}{\includegraphics{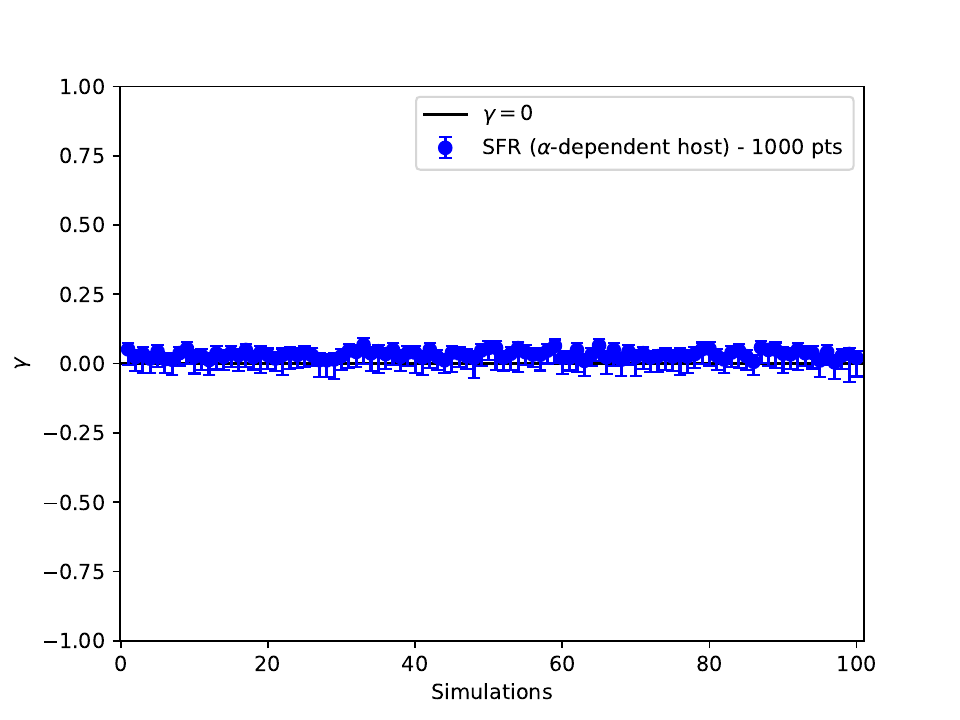}}\\
\end{center}
\caption{The same as in the previous figure, considering N = 1000. }  
\label{lowexp1}
\end{figure*}

For completeness, we present below the best-fit results of our simulations for $\gamma$ assuming $N = 500$ (Fig. \ref{lowexp}) and $N = 1000$ (Fig. \ref{lowexp1}).

%%%%%%%%%%%%%%%%%%%%%%%%%%%%%%%%%%%%%%%%%%%%%%%%%%

\end{document}